\documentstyle[12pt] {article}
\def\zid{1\kern-0.36em\llap~1}

\newcommand{\beq}{\begin{equation}}
\newcommand{\ber}{\begin{eqnarray}}
\newcommand{\eeq}{\end{equation}}
\newcommand{\eer}{\end{eqnarray}}

\begin{document}

\begin{titlepage}
\rightline{SUNY BING 5/21/98}
\vspace{1mm}
\begin{center}
{\bf THE STATE-VECTOR SPACE FOR TWO-MODE PARABOSONS\newline
and CHARGED PARABOSE COHERENT STATES}\\
\vspace{2mm}
Sicong Jing\footnote{On leave from:  Department of
Modern Physics, University of Science and Technology of 
China, Hefei, 230026, P.R. China.} and 
Charles A. Nelson\footnote{Electronic address: cnelson @ 
bingvmb.cc.binghamton.edu }\\
{\it Department of Physics, State University of New York at 
Binghamton\\
Binghamton, N.Y. 13902-6016}\\[2mm]
\end{center}


\begin{abstract} 

The structure of the state-vector space for the two-mode 
parabose system is investigated and a complete set of
 state-vectors is constructed.  The basis vectors
are orthonormal in order $p=2$. In order $p=2$,
conserved-charge parabose coherent states
are 
constructed and an explicit completeness relation
is obtained.  
\end{abstract}

\end{titlepage}

\section{Introduction}

A fundamental unresolved question in physics is whether all 
particles of
nature are necessarily either bosons or fermions. Theoretical 
investigations 
of other possibilities in local, relativistic quantum field 
theory 
show that there
may exist more general particle statistics [1-2]. There may 
exist three
types of statistics for identical particles: the parabose and 
parafermi
statistics for which the number of particles in an 
antisymmetric or a
symmetric state, respectively, cannot exceed a given integer 
$p$ called the
order of the parastatistics, and for two space dimensions,   
infinite statistics based on the braid group.

Knowledge of the structure of the state-vector space for a 
quantum theory is
essential. For example, many successful applications of the 
ordinary boson
and fermion descriptions in various fields of physics are 
based on the full
knowledge about the structure of the ordinary bose and fermi 
state-vector
spaces. During the period of early interest in 
parastatistics, the explicit
structure of the state-vector space for a single-mode of 
parabosons, and of
parafermions, was determined. The associated coherent states 
for a single
parabose mode was also constructed [3]. Because there is no 
simple
commutative or anticommutative bilinear relations between 
single paraparticles  
belonging to
different degrees of freedom, knowledge about the structure 
of the
state-vector space for the case of more than one mode of 
paraparticles
remains very limited. One knows that the space is 
spanned by a
state vector of the form ${\cal M}(a_k^{\dagger 
},a_l^{\dagger 
},\cdots )|0>,$ where
${\cal M}$
denotes an arbitrary monomial in the parabose creation 
operators $%
a_k^{\dagger },a_l^{\dagger },\cdots ,$ and $|0>$ is the
unique vacuum state.  
While this $n$-paraparticle state vector can be written[4] as 
a 
so-called
``state-vector of the standard form'', one still doesn't know 
the explicit
form of a complete set of basis vectors for such systems. 
Consequently,
parabose coherent states for more than the one mode case have 
not been
constructed. 

In this paper, we investigate the simplest non-trivial 
case---the structure
of the state space for the two-mode parabose system. While 
the fundamental
parabose commutation relations are trilinear, 
\begin{equation}
\begin{array}{c}
\lbrack a_k,\{a_l^{\dagger },a_m\}]=2\delta 
_{kl}a_m, \; [a_k,\{a_l^{\dagger
},a_m^{\dagger }\}]=2\delta _{kl}a_m^{\dagger }+2\delta 
_{km}a_l^{\dagger },
\\ 
\lbrack a_k,\{a_l,a_m\}]=0, \hspace{2pc} (k,l,m=1,2)
\end{array}
\end{equation}
it does nevertheless follow that there are some simple 
commutation relations
between ``A'' paraboson operators and the ``B'' paraboson
operators: letting 
$a\equiv a_1,$ $b\equiv a_2$, 
\begin{equation}
[a,b^2]=0, \; [b,a^2]=0, \; [a^{\dagger },b^2]=0, \; 
[b^{\dagger
},a^2]=0
\end{equation}
plus the hermitian conjugate relations.

In Sec. 2 we use these relations to construct an explict 
complete set of
state-vectors for the two-mode parabose system. Then  for 
$p=2$ order
parabosons, in Sec. 3 we show that these state-vectors are 
also orthogonal
and thereby obtain a complete, orthonormal set of basis 
vectors. In Sec. 4,
we use these basis vectors to construct the conserved-charge 
parabose
coherent states and show that they satisfy an explicit 
completeness relation.

\section{The Complete Set of State-Vectors}

We assume there is a unique vacuum state $|0>$ satisfying 
\begin{equation}
a_k|0>=0, \; a_ka_l^{\dagger }|0>=p\delta _{kl}|0>
\end{equation}
and consider a state with a total number $N$ of parabosons. 
Without loss of
generality, in this state there are $n$ parabosons $A$ and 
$m$ parabosons $B$%
, with $0\leq n,m\leq N$ and $n+m=N$.  We denote each 
partition by $(n,m)$.  For a given $N$, the number of its 
partitions is $N+1$.

{\it Theorem:} The dimension of the subspace $(n,m)$ is $\min 
(n,m)+1$.

To prove this, without loss of generality we assume that 
$n\geq m$. Counting
the number of states in the subspace $(n,m)$ is equal to 
counting the number
of ways of arranging $n$ particles $A$ and $m$ particles $B$ 
in $n+m=N$
boxes. While $A$ and $B$ cannot be freely interchanged, by 
(2) two adjacent $%
A$ particles can be freely interchanged with a single $B$ 
particle, and vice
versa. For $m=$even, we first assign the $m$ particles $B$ in 
the last $m$
boxes, with the other boxes occupied by the $n$ particles 
$A$. We call this
state 1, and denote it $AA\cdots ABB\cdots B$. Without 
separating the $B$
particles, no new state arises. So we next separate one $B$ 
particle and
put it in the last box, with one $A$ in the next box, and 
then put the
remaining $B$ particles in the next $m-1$ boxes. This is a 
new state (state
2) denoted by $A\cdots AB\cdots BAB$. Again, 
repositioning the
group of $m-1$ $B$ particles dosn't produce a state different 
from states 1
or 2. Next we put $ABA$ in the last $3$ boxes, and the $m-1$ 
group of $B$
particles in the next  $m-1$ boxes, followed by the $n-2$ $A$ 
particles.
This is a new state (state 3) denoted by $A\cdots AB\cdots 
B\cdots BABA$. 
Again, repositioning the group of $m-1$ $B$ particles dosn't 
produce a new
state. The next step is to separate $3$ $B$ particles on the 
right, and 
insert $A$ particles to 
keep these $3$ $B$
particles separate, while the remaining $m-3$ $B$ 
particles are kept grouped together. Proceeding as before, 
this again 
gives two new states: 
$A\cdots AB\cdots
BABABAB$ and $A\cdots AB\cdots BABABABABA.$ Continuing this 
process, each
time we separate two additional $B$ particles and obtain two 
different
states with different right endings. This procedure ends when 
it produces two states with all the $B$ particles non-
adjacent.  Thereby, for $m=$even, 
we obtain $%
2\left[ \frac m2\right] +1=m+1$ different 
states\footnote{For $x\geq 0$, square brackets $[x]$ 
denote 
the integer part of $x$.} . 
Similarly for  $m=$odd, we construct $2\left[ 
\frac{m+1}2\right] =m+1$
different states.

This theorem reflects a major difference between the 
paraboson system and
the ordinary boson system. In the ordinary boson case, there 
is only one
state in the subspace $(n,m)$ , i.e. $(a^{\dagger 
})^n(b^{\dagger })^m|0>$,
versus $m+1$ different states ( if $n\geq m$ ) in the 
paraboson system. It
also follows from this theorem, that the dimension of the 
state-vector space
with a total of $N$ paraboson particles $A$ and $B$ is 
$([\frac N2]+1)([%
\frac{N-1}2]+2)$, instead of  the dimension $N+1$ for the 
ordinary two-mode
boson case.

In summary, for all $n$ and $m$ values, we can write the 
state vector of $n$ parabosons 
$A$ and $m$
parabosons $B$ as 
\begin{equation}
|n,m;i>=\frac 1{\sqrt{N_{n,m}^i}}(a^{\dagger })^{n-
i+S}(b^{\dagger })^{m-2[%
\frac{i-S}2]}(a^{\dagger }b^{\dagger })^{2 [\frac{i-
S}2]}(a^{\dagger })^{i-S-2[%
\frac{i-S}2]}|0> 
\end{equation}
where $N_{n,m}^i$ is the normalization constant, and 
$$
S={ \frac 12 }(1-(-)^m), \; 1\leq i\leq \min (n,m)+1. 
$$

It is useful to note that the appearance of a new index $i$ 
is a
characteristic feature of para-systems with more than one 
mode. This occurs
because of the intrinsic degeneracy of the many-mode 
parasystem; i.e. the
quantum numbers $n$ and $m$ do not suffice to completely 
specify the quantum
states since $ab \neq ba$.  Since the proof is constructive, 
there is no 
possibility to build
any other states, and 
so the set of state vectors%
$$
\{|n,m;i>| \; n,m=0,1,\cdots ; \; 1\leq i\leq \min (n,m)+1\} 
$$
is complete.

These state vectors are orthogonal between different $(n,m)$ 
subspaces, but
in general the Gram-Schmidt orthogonalization method or some 
other 
procedure must
be used for different $i$ and $i^{\prime }$ states in such a 
subspace.
However, this latter step is not necessary for order 
$p=2$ as we next show.

\section{The Order  $p=2$ Case}

Inspection of the structure of the state-vector given in (4) 
shows that
there are two distinct orderings of the parabosons $A$ and 
$B$ in the $(n,m)$
subspace: \newline
(i) Type I: $(a^{\dagger })^{n-2j}(b^{\dagger })^{m-
2j}(a^{\dagger
}b^{\dagger })^{2j}|0>$, and  \newline
(ii) Type II: $(a^{\dagger
})^{n-2j-1}(b^{\dagger })^{m-2j-1}(b^{\dagger }a^{\dagger 
})^{2j+1}|0>$,
where $j=[\frac{i-S}2]$. Type I(II) respectively corresponds 
to $i-S$ being
even (odd), where $S$ is defined after (4).

Neglecting the normalization factors $N_{n,m}^i$and 
$N_{n,m}^{i^{\prime }}$,
we have when both state vectors are \newline type I
\begin{equation}
<n,m;i^{\prime }|n,m;i>=<0|(ba)^{2j^{\prime }}b^{m-2j^{\prime
}}a^{n-2j^{\prime }}(a^{\dagger })^{n-2j}(b^{\dagger })^{m-
2j}(a^{\dagger
}b^{\dagger })^{2j}|0> 
\end{equation}
where $j^{\prime }=[\frac{i^{\prime }-S}2]$.  With no loss in 
generality 
we assume $%
j>j^{\prime }$. Using the algebraic relations 
\begin{equation}
a(a^{\dagger })^{2n}=2n(a^{\dagger })^{2n-1}+(a^{\dagger
})^{2n}a, \; b(b^{\dagger })^{2n}=2n(b^{\dagger })^{2n-
1}+(b^{\dagger })^{2n}b 
\end{equation}
and (only true for $p=2$) 
\begin{equation}
ba^{\dagger }b^{\dagger }=b^{\dagger }a^{\dagger 
}b, \; ab^{\dagger }a^{\dagger
}=a^{\dagger }b^{\dagger }a, \; bb^{\dagger }a^{\dagger 
}=a^{\dagger }b^{\dagger
}b+2a^{\dagger }, \; aa^{\dagger }b^{\dagger }=b^{\dagger 
}a^{\dagger 
}a+2b^{\dagger } 
\end{equation}
we get\footnote{Note $(2k)!!=2k(2k-2)(2k-4)\cdots 2$, 
$(2k+1)!!=(2k+1)(2k-1)(2k-3)\cdots 1$ .}
\begin{displaymath}
<n,m;i^{\prime }|n,m;i>=  
\end{displaymath}
\begin{equation}
 \left\{ \begin{array}{ll}
\frac{(m-2j)!!(m+2j)!!}{(4j)!!}<0|(ba)^{2j^{\prime }}a^{n-
2j^{\prime
}}(a^{\dagger })^{n-2j}(b)^{2(j-j^{\prime })}(a^{\dagger 
}b^{\dagger
})^2|0>=0 & (m \; \mbox{even} ) \\
\frac{(m-1-
2j)!!(m+1+2j)!!}{(4j)!!}%
<0|(ba)^{2j^{\prime }}ba^{n-2j^{\prime }}(a^{\dagger
})^{n-2j}(b)^{2(j-j^{\prime })-1}(a^{\dagger }b^{\dagger 
})^2|0>=0 & (m \; \mbox{odd} )  \\
\end{array}
\right.
\end{equation}
\newline 
In $(8)$ we used the fact that $bb^{\dagger }(a^{\dagger 
}b^{\dagger
})^n|0>=2(n+1)(a^{\dagger }b^{\dagger })^n|0>$ and \newline  
$b(a^{\dagger
}b^{\dagger })^n|0>=(b^{\dagger }a^{\dagger })^nb|0>=0$.

Next we consider the case of both the state vectors being 
type II: 
\begin{equation}
<n,m;i^{\prime }|n,m;i>=<0|(ab)^{2j^{\prime }+1}b^{m-
2j^{\prime
}-1}a^{n-2j^{\prime }-1}(a^{\dagger })^{n-2j-1}(b^{\dagger
})^{m-2j-1}(b^{\dagger }a^{\dagger })^{2j+1}|0>
\end{equation}
If $m$ is odd, this overlap (9) can be written as 
\begin{displaymath}
<0|(ab)^{2j^{\prime }+1}b^{m-2j^{\prime }-1}(b^{\dagger
})^{m-2j-1}a^{n-2j^{\prime }-1}(a^{\dagger })^{n-2j-
1}(b^{\dagger
}a^{\dagger })^{2j+1}|0>= 
\end{displaymath}
\begin{equation}
\left\{ \begin{array}{ll}
\frac{(n-2-2j)!!(n+2+2j)!!}{(4j+2)!!}<0|(ab)^{2j^{\prime 
}+1}b^{m-2j^{\prime
}-1}(b^{\dagger })^{m-2j-1}a^{2(j-j^{\prime })}(b^{\dagger 
}a^{\dagger
})^{2j+1}|0>=0  & (n \; \mbox{even} ) \\
\frac{(n-1-
2j)!!(n+1+2j)!!}{(4j+2)!!%
}<0|(ab)^{2j^{\prime }+1}b^{m-2j^{\prime }-1}(b^{\dagger
})^{m-2j-1}a^{2(j-j^{\prime })}(b^{\dagger }a^{\dagger 
})^{2j+1}|0>=0  & (n \; \mbox{odd} )  \\
\end{array}
\right. 
\end{equation} 
\newline 
If $m$ is even, the overlap (9) can be written as 
\begin{displaymath}
<0|(ab)^{2j^{\prime }+1}b^{m-2j^{\prime }-1}(b^{\dagger
})^{m-2j}a^{n-2j^{\prime }-1}(a^{\dagger })^{n-2j}(b^{\dagger 
}a^{\dagger
})^{2j}|0>= 
\end{displaymath}
\begin{equation}
\left\{ \begin{array}{ll}
\frac{(n-2j)!!(n+2j)!!}{(4j)!!}<0|(ab)^{2j^{\prime }+1}b^{m-
2j^{\prime
}-1}(b^{\dagger })^{m-2j}a^{2(j-j^{\prime })-1}(b^{\dagger 
}a^{\dagger
})^{2j}|0>=0  & (n \; \mbox{even} ) \\
\frac{(n-1-
2j)!!(n+1+2j)!!}{(4j)!!}%
<0|(ab)^{2j^{\prime }+1}b^{m-2j^{\prime }-1}(b^{\dagger
})^{m-2j}a^{2(j-j^{\prime })-1}(b^{\dagger }a^{\dagger 
})^{2j}|0>=0  & (n \; \mbox{odd} )  \\
\end{array} 
\right.
\end{equation}
\newline 
Finally, we consider the overlap of one state vector 
belonging to the type I
and another state vector belonging to the type II: 
\begin{displaymath} 
<n,m;i^{\prime }|n,m;i>=<0|(ab)^{2j^{\prime }+1}b^{m-
2j^{\prime
}-1}a^{n-2j^{\prime }-1}(a^{\dagger })^{n-2j}(b^{\dagger
})^{m-2j}(a^{\dagger }b^{\dagger })^{2j}|0>= 
\end{displaymath}
\begin{equation}
\left\{ \begin{array}{ll}
\frac{(m-2j)!!(m+2j)!!}{(4j)!!}<0|(ab)^{2j^{\prime }}a^{n-
2j^{\prime
}}(a^{\dagger })^{n-2j}b^{2(j-j^{\prime })}(a^{\dagger 
}b^{\dagger
})^{2j}|0>=0  & (m \; \mbox{even} ) \\ 
\frac{(m-1-
2j)!!(m+1+2j)!!}{(4j)!!}%
<0|(ab)^{2j^{\prime }+1}a^{n-2j^{\prime }-1}(a^{\dagger
})^{n-2j}b^{2(j-j^{\prime })-1}(a^{\dagger }b^{\dagger 
})^{2j}|0>=0  & (m \; \mbox{odd} )  \\
\end{array} 
\right. 
\end{equation}
\newline 
Thus, 
\begin{equation}
<n,m;i^{\prime }|n,m;i>=0 \; \mbox{for} \; i\neq i^{\prime }
\end{equation}
which completes the proof of orthogonality for $p=2$ for the 
state vectors
given by $(4)$.

The normalization constant $N_{n,m}^i$ for the state vector 
$|n,m;i>$ easily
follows from the algebraic relations $(7)$,
\begin{equation}
(N_{n,m}^i)^2=2^{n+m}[\frac{m+i}2]![\frac{n+1-
i}2]![\frac{m+i}2]![\frac{m+1-i%
}2]!
\end{equation}

When the annihilation operators $a$ and $b$ act on this set 
of basis
vectors, one finds 
\begin{equation}
a|n,m;i>=
\left\{ \begin{array}{ll}
\sqrt{2[\frac{n+i}2]}|n-1,m;i>  & \mbox{if} \; (n+i) \; 
\mbox{even}  \\
\sqrt{2[\frac{%
n+i-1}2]}|n-1,m;i> & \mbox{if} \; (n+i) \; \mbox{odd}  \\
\end{array}
\right.
\end{equation}
and
\begin{equation}
b|n,m;i>=
\left\{ \begin{array}{ll}
\sqrt{2[\frac{m+i-1}2]}|n,m-1;i+1> & \mbox{if} \; (n+i) \; 
\mbox{even}  \\
\sqrt{2[\frac{%
m+i}2]}|n,m-1;i-1> & \mbox{if} \; (n+i) \; \mbox{odd}  \\
\end{array}
\right.
\end{equation}
When we use $(16)$ for  $i=1$,  we identify $|n,m-1;i=0>$
with $%
|n,m-1;i=1>$ since the construction of (4) does not include 
$i=0$.  For instance since $|2,2;1>=\frac 14 
(a^{\dagger})^2 (b^{\dagger})^2 |0>$
and $|2,1;1>=\frac 1{2\sqrt{2}} (a^{\dagger})^2 b^{\dagger 
}|0>$, 
we have $%
b|2,2;1>=
\frac 12 (a^{\dagger})^2 b^{\dagger
}|0>=\sqrt{2}|2,1;1>$.

Similarly, when the creation operators act on the basis 
vectors,  
\begin{equation}
a^{\dagger }|n,m;i>=
\left\{ \begin{array}{ll}
\sqrt{2[\frac{n+2-i}2]}|n+1,m;i> & \mbox{if} \; (n+i) \; 
\mbox{even}  \\
\sqrt{2[\frac{%
n+1+i}2]}|n+1,m;i> & \mbox{if} \; (n+i)  \; \mbox{odd}  \\
\end{array}
\right. 
\end{equation}
and
\begin{equation}
b^{\dagger }|n,m;i>=
\left\{ \begin{array}{ll}
\sqrt{2[\frac{m+2+i}2]}|n,m+1;i+1> & \mbox{if} \; (n+i) \; 
\mbox{even}  \\
\sqrt{2[\frac{%
m+3-i}2]}|n,m+1;i-1> & \mbox{if} \; (n+i) \; \mbox{odd}  \\
\end{array}
\right. 
\end{equation}
In (18), for $i=1$ we identify $|n,m+1;i=0>$ with 
$|n,m+1,i=1>$.

The parabose number operators $N_a$ and $N_b$ for $p=2$ order 
are
respectively defined by 
\begin{equation}
N_a=\frac 12\{a^{\dagger },a\}-1, \; N_b=\frac 12\{b^{\dagger 
},b\}-1
\end{equation}
From (15)-(18)
\begin{equation}
N_a|n,m;i>=n|n,m;i>, \; N_b|n,m;i>=m|n,m;i>,
\end{equation}
so the state vectors $|n,m;i>$ are common eigenvectors of 
$N_a,N_b$, and, thus, are two-mode parabose number
states.

\section{Conserved-Charge Parabose Coherent States \newline 
for Order $p=2$}

As an application of the complete set of orthonormal state 
vectors for the
two-mode parabose system for $p=2$ order, we construct the 
associated 
conserved-charge parabose coherent states. In  physics 
applications of
coherent state techniques it is normally necessary to make 
various
approximations, but it also often remains important to 
maintain the
conservation of an Abelian charge.

Using the above number operators $N_a$ and $N_b$, we define a 
hermitian,
charge operator by 
\begin{equation}
Q=N_a-N_b
\end{equation}
so each of the $A$ quanta possesses a charge ``$+1$'' and 
each 
of the $B$
quanta a charge ``$-1$''. Since $Q$ does not commute with $a$ 
or $b$, we
cannot require that the coherent state be simultaneously an 
eigenstate of $Q$
and the annihilation operators $a$ and/or $b$. Since 
\begin{equation}
[Q,ab]=0, \; [Q,ba]=0, \; [ab,ba]=0,
\end{equation}
we define the conserved-charge parabose coherent state 
$|q,z,z^{\prime }>$
by the requirements that
\begin{equation}
Q|q,z,z^{\prime }>=q|q,z,z^{\prime }>, \; ab|q,z,z^{\prime 
}>=z|q,z,z^{\prime
}>, \; ba|q,z,z^{\prime }>=z^{\prime }|q,z,z^{\prime }>
\end{equation}
Here for parabosons since  $ab\neq ba$, we introduce two 
complex numbers $z$
and $z^{\prime }$, unlike for ordinary bosons ($p=1$) where 
only one $z$ was
needed[5].

To obtain an explicit expression for these coherent states, 
we consider the
expansion
\begin{equation}
|q,z,z^{\prime }>=\sum_{n,m=0}^\infty \sum_{i=1}^{\min
(n,m)+1}c_{n,m}^i|n,m;i>
\end{equation}
with the $c_{n,m}^i$ expansion coefficents to be determined. 
Since $%
|q,z,z^{\prime }>$ is an eigenstate of $Q$, for $q\geq 0$ 
\begin{equation}
|q,z,z^{\prime }>=\sum_{m=0}^\infty 
\sum_{i=1}^{m+1}c_{q+m,m}^i|q+m,m;i>
\end{equation}
Substituting this expression into the remaining two 
eigen-equations in (23) and using (15)-(16), we obtain
\begin{equation}
c_{q+m,m}^i(z,z^{\prime })=c_{q,0}^1\frac{\sqrt{[\frac 
q2]![\frac{q+1}2]!}%
(z)^{[\frac{m-(-)^{q+m+i}i}2+\frac{1-(-)^q}4]}(z^{\prime 
})^{[\frac{%
m+(-)^{q+m+i}i}2+\frac{1+(-
)^q}4]}}{2^m\sqrt{[\frac{m+i}2]![\frac{q+m+i}2]![%
\frac{m+1-i}2]![\frac{q+m+1-i}2]!}}
\end{equation}
Thus, the charged parabose coherent state for $q\geq 0$ is
\begin{equation}
|q,z,z^{\prime }>=N_q(z,z^{\prime })\sum_{m=0}^\infty 
\sum_{i=1}^{m+1}\frac{%
(z)^{[\frac{m-(-)^{q+m+i}i}2+\frac{1-(-)^q}4]}(z^{\prime 
})^{[\frac{%
m+(-)^{q+m+i}i}2+\frac{1+(-
)^q}4]}}{2^m\sqrt{[\frac{m+i}2]![\frac{q+m+i}2]![%
\frac{m+1-i}2]![\frac{q+m+1-i}2]!}}|q+m,m;i>
\end{equation}
with the normalization constant 
\ber
(N_q)^{-2}=\sum_{m=0}^\infty \sum_{i=1}^{m+1}
\frac{|z|^{2[\frac{m-(-)^{q+m+i}i}2+\frac{1-(-
)^q}4]}|z^{\prime }|^{2[\frac{%
m+(-)^{q+m+i}i}2+\frac{1+(-
)^q}4]}}{2^{2m}[\frac{m+i}2]![\frac{q+m+i}2]![%
\frac{m+1-i}2]![\frac{q+m+1-i}2]!} \nonumber \\
 =(i\frac{|z|}2)^{-
[\frac q2]}J_{[\frac
q2]}(i|z|)(i\frac{|z^{\prime }|}2)^{-
[\frac{q+1}2]}J_{[\frac{q+1}%
2]}(i|z^{\prime }|)
\eer
where $J_n$ is a Bessel function of order $n$.

When $q<0$, the construction proceeds similarly: The charged 
parabose
coherent state is 
\begin{equation}
|q,z,z^{\prime }>=\sum_{m=0}^\infty 
\sum_{i=1}^{m+1}c_{m,|q|+m}^i|m,|q|+m;i>
\end{equation}
where
\begin{equation}
c_{m,|q|+m}^i(z,z^{\prime })=c_{0,q}^1\frac{\sqrt{[\frac 
{|q|}2]![\frac{|q|+1}2]!}%
(z)^{[\frac{m-(-)^{m+i}i}2]}(z^{\prime })^{[\frac{m+(-
)^{m+i}i+1}2]}}{2^m%
\sqrt{[\frac{m+i}2]![\frac{|q|+m+i}2]![\frac{m+1-
i}2]![\frac{|q|+m+1-i}2]!}}
\end{equation}
The charged parabose coherent state for $q<0$ is
\begin{equation}
|q,z,z^{\prime }>=N_q(z,z^{\prime })\sum_{m=0}^\infty 
\sum_{i=1}^{m+1}\frac{%
(z)^{[\frac{m-(-)^{m+i}i}2]}(z^{\prime })^{[\frac{m+(-
)^{m+i}i+1}2]}}{2^m%
\sqrt{[\frac{m+i}2]![\frac{|q|+m+i}2]![\frac{m+1-
i}2]![\frac{|q|+m+1-i}2]!}}%
|m,|q|+m;i>
\end{equation}
with the normalization constant for $q<0$%
\ber
(N_q)^{-2}=\sum_{m=0}^\infty \sum_{i=1}^{m+1}
\frac{|z|^{2[\frac{m-(-)^{m+i}i}2]}|z^{\prime 
}|^{2[\frac{m+(-)^{m+i}i+1}2]}%
}{2^{2m}[\frac{m+i}2]![\frac{|q|+m+i}2]![\frac{m+1-
i}2]![\frac{|q|+m+1-i}2]!} \nonumber \\ 
=(i\frac{|z|}2)^{-
[\frac{|q|+1}2]}J_{[\frac{|q|+1}2]}(i|z|)(i\frac{%
|z^{\prime }|}2)^{-
[\frac{|q|}2]}J_{[\frac{|q|}2]}(i|z^{\prime }|)
\eer

The inner product of two non-negatively charged parabose 
coherent states is $%
(q,q^{\prime }\geq 0)$ 
\begin{equation}
<q,z,z^{\prime }|q^{\prime },w,w^{\prime }>=\delta 
_{q,q^{\prime }}\frac{%
( \frac{i}2
 w^{*}z)^{-[\frac q2]}J_{[\frac q2]}(iw^{*}z)( \frac{i}2
 (w^{\prime 
})^{*}z^{\prime
})^{-[\frac{q+1}2]}J_{[\frac{q+1}2]}(i(w^{\prime 
})^{*}z^{\prime })}{%
 N_q(z,z^{\prime }) N_{q^{\prime }}  
(w,w^{\prime })}
\end{equation}
and for $(q,q^{\prime }<0)$%
\begin{equation}
<q,z,z^{\prime }|q^{\prime },w,w^{\prime }>=\delta 
_{q,q^{\prime }}\frac{%
( \frac{i}2 
 w^{*}z)^{-
[\frac{|q|+1}2]}J_{[\frac{|q|+1}2]}(iw^{*}z)( \frac{i}2
 (w^{\prime
})^{*}z^{\prime })^{-
[\frac{|q|}2]}J_{[\frac{|q|}2]}(i(w^{\prime
})^{*}z^{\prime })} {%
 N_q(z,z^{\prime }) N_{q^{\prime }}  
(w,w^{\prime })}
\end{equation}
If $q\geq 0$ and $q^{\prime }<0$, the inner product vanishes. 
Therefore, the
charged parabose coherent states with different charges are 
orthogonal, but
for the same $q$-sector, the charged parabose coherent states 
are not
orthogonal. Consequently, for the same $q$-sector the charged 
parabose
coherent states are linearly dependent and overcomplete.

These charged parabose coherent states satisfy the 
completeness relation
\begin{equation}
\sum_{q=-\infty }^\infty \int \frac{d^2zd^2z^{\prime }}{\pi 
^2}\Phi
_q(z,z^{\prime })|q,z,z^{\prime }><q,z,z^{\prime }|=I
\end{equation}
where $d^2z=rdrd\theta $, $d^2z^{\prime }=r^{\prime 
}dr^{\prime }d\theta
^{\prime }$, and
\begin{equation}
\Phi _q(z,z^{\prime })= 
\left\{ \begin{array}{ll}
\frac 14(-i)^{[\frac 
q2]+[\frac{q+1}%
2]}J_{[\frac q2]}(i|z|)K_{_{[\frac 
q2]}}(|z|)J_{_{[\frac{q+1}%
2]}}(i|z^{\prime }|)K_{_{[\frac{q+1}2]}}(|z^{\prime }|) & 
\mbox{for} \; q\geq 0 \\
\frac 14(-
i)^{[\frac{|q|}2]+[\frac{|q|+1}2]}J_{[\frac{|q|+1}2]}(i|z|)K_
{_{[%
\frac{|q|+1}2]}}(|z|)J_{_{[\frac{|q|}2]}}(i|z^{\prime 
}|)K_{_{[\frac{|q|}%
2]}}(|z^{\prime }|)  & 
\mbox{for} \; q< 0 \\
\end{array}
\right.
\end{equation}
with $K_n(x)=\frac \pi 2i\exp (\frac{in\pi 
}2)(J_n(ix)+iN_n(ix))$ a
modified Bessel function. This result follows since by the 
integration formula [6]%
\begin{equation}
\int_o^\infty dr \; r^\mu K_\nu (ar)=2^{\mu -1}a^{-\mu -
1}\Gamma 
(\frac{\mu +\nu
+1}2)\Gamma (\frac{\mu -\nu +1}2), \; (Re(\mu \pm \nu 
)>0, \; Re(a)>0), 
\end{equation}
we find  
\begin{eqnarray}
\sum_{q=-\infty }^\infty \int 
\frac{d^2zd^2z^{\prime }}{\pi ^2}\Phi _q(z,z^{\prime 
})|q,z,z^{\prime
}><q,z,z^{\prime }| 
=\sum_{q=0}^\infty \sum_{m=0}^\infty
\sum_{i=1}^{m+1}|q+m,m;i><q+m,m;i|\nonumber \\ 
+ \sum_{-q=1}^\infty 
\sum_{m=0}^\infty
\sum_{i=1}^{m+1}|m,|q|+m;i><m,|q|+m;i|  
=\sum_{n,m=0}^\infty \sum_{i=1}^{\min 
(n,m)+1}|n,m;i><n,m;i|=I
\end{eqnarray} 

In summary, in this paper we construct a complete set of 
basis vectors for
the two-mode paraboson system. In order $p=2$, the basis 
vectors are
orthonormal and we construct the associated 
conserved-charge parabose 
coherent states. The latter are orthogonal between different 
$q$-sectors and are overcomplete within each $q$-sector. It 
is important to generalize these constructions to more
than the two-mode system and to $p>2$.

This work was partially supported by the National Natural 
Science Foundation of China and by U.S. Dept. of Energy 
Contract No. DE-FG 02-96ER40291.

\end{document}